\documentclass[10pt,conference]{IEEEtran}
\IEEEoverridecommandlockouts
\IEEEoverridecommandlockouts
\usepackage{cite}
\usepackage{amsmath,amssymb,amsfonts}
\usepackage{algorithmic}
\usepackage{graphicx}
\usepackage{textcomp}
\usepackage{xcolor}
\def\BibTeX{{\rm B\kern-.05em{\sc i\kern-.025em b}\kern-.08em
    T\kern-.1667em\lower.7ex\hbox{E}\kern-.125emX}}
\newcommand\figref{Fig.~\ref}
\usepackage{microtype}

\usepackage[font = footnotesize]{caption} 
\begin{document}
\title{Clutter Suppression in Bistatic ISAC with Joint Angle and Doppler Estimation}
\author{
M. Ertug Pihtili\textsuperscript{\dag}, %~\IEEEmembership{Graduate Student Member,~IEEE}, 
Julia Equi\textsuperscript{\ddag}, %~\IEEEmembership{Member,~IEEE}, 
Ossi Kaltiokallio\textsuperscript{\dag}, %~\IEEEmembership{Member,~IEEE},
Jukka Talvitie\textsuperscript{\dag}, %~\IEEEmembership{Member,~IEEE}, 
Elena Simona Lohan\textsuperscript{\dag}, %~\IEEEmembership{Senior Member,~IEEE},  
Ertugrul Basar\textsuperscript{\dag}, \\%~\IEEEmembership{Fellow,~IEEE}, \\
and Mikko Valkama\textsuperscript{\dag} \\%~\IEEEmembership{Fellow,~IEEE} \\
\textsuperscript{\dag}Tampere Wireless Research Center, Electrical Engineering Unit, Tampere University, Tampere, Finland \\
\textsuperscript{\ddag}Ericsson Research, Helsinki, Finland \\
%Email: ertug.pihtili@tuni.fi \\
%Email: julia.equi@ericsson.com , ossi.kaltiokallio, jukka.talvitie, elena-simona.lohan, ertugrul.basar, mikko.valkama\
}
\maketitle
\begin{abstract}
\textls[-1]{The coexistence of radar and communications in wireless systems marks a paradigm shift for the sixth-generation (6G) networks. As 6G systems are expected to operate at higher frequencies and employ larger antenna arrays than fifth-generation (5G) systems, they can also enable more accurate sensing capabilities. To this end, the integrated sensing and communication (ISAC) paradigm aims to unify the physical and radio frequency (RF) domains by introducing the sensing functionality into the communication network. However, the clutter poses a challenge, as it can significantly degrade the sensing accuracy in ISAC systems. This paper presents a novel two-dimensional root multiple signal classification (2D-rootMUSIC)-based algorithm for static background clutter suppression. Computer simulation results indicate that the proposed method effectively mitigates the strong background clutter, yields accurate parameter estimation performance, and offers a notable improvement in the signal-to-clutter-and-noise ratio (SCNR), while outperforming the prior-art benchmark methods.}
\end{abstract}


\vspace{-2mm}
\section{Introduction}
\vspace{-1mm}
\textls[-6]{Next-generation networks opt for larger bandwidths, increased antenna arrays, and higher frequencies to meet the demands of sixth-generation (6G) systems. These advancements enable the convergence of radar and communication technologies, fostering the development of perceptive networks that incorporate sensing functionalities within communication systems, commonly referred to as integrated sensing and communication (ISAC) \cite{Gonzalez2024}. As a result, 6G systems are expected to support a wide range of use cases by leveraging the dual functionalities of ISAC infrastructures. A recent European Telecommunications Standards Institute (ETSI) report outlines 18 distinct use cases enabled by the network's sensing functionality, ranging from human motion recognition to sensing-assisted communications \cite{etsi2025isac}. Furthermore, utilizing a common hardware platform and shared waveform for both functionalities presents a promising approach to building smarter networks, while allowing for coordination and integration gains in 6G ISAC systems due to the similar requirements of sensing and communication tasks \cite{Wei2022}.}

\textls[-6]{In this respect, orthogonal frequency division multiplexing (OFDM) emerges as a suitable waveform, as it can serve as a radar probing signal for environmental sensing while simultaneously transmitting information to communication users. Rather than using conventional radar waveforms, a modulation symbol-based approach such as quadrature amplitude modulation (QAM) or phase shift keying (PSK) through the OFDM waveform enables the dual functionality of ISAC systems effectively \cite{Sturm2011}. In such systems, the fast Fourier transform (FFT) and inverse FFT (IFFT) are applied to the time and frequency resources of the OFDM signal, specifically across the symbol and subcarrier domains, for Doppler and range estimation, respectively. In addition, adopting larger bandwidths and higher frequencies promotes the transition from frequency range 1 (FR1) to frequency range 2 (FR2), also known as the millimeter-wave (mmWave) band. Consequently, in multiple-input multiple-output (MIMO) OFDM systems, the use of mmWave bands enhances parameter estimation accuracy in sensing and localization tasks due to improved resolution, although it introduces increased complexity and necessitates larger antenna array sizes \cite{Gonzalez2024}.}

\textls[-6]{In ISAC, sensing refers to the detection of an object and the estimation of its parameters by analyzing variations in radio frequency (RF) signals as they propagate from the transmitter (Tx) to the receiver (Rx) within the sensing area. The sensing operations can be realized through monostatic, bistatic, and multistatic modes in ISAC architectures. A monostatic mode exploits the same transceiver for both transmission and reception, yet it suffers from a self-interference problem due to the full-duplex signaling scheme. To address this, the bistatic mode can be a potential solution, where the sensing Tx and Rx are spatially separated to mitigate self-interference, and multistatic sensing extends the bistatic mode by employing several Tx or Rx units \cite{Gonzalez2024}. To facilitate sensing in ISAC systems, the authors in \cite{Chen2023} proposed a multiple signal classification (MUSIC)-based ISAC architecture to separately estimate delay, Doppler, and angle of arrival (AoA). Furthermore, the joint estimation of these parameters through FFT and IFFT operations using a radar data cube was presented in \cite{Xiao2024}. The authors in \cite{Pucci2022} investigated the performance of a bistatic sensing system by considering the fifth-generation new radio (5G NR) system setup.}

\textls[-6]{In this context, the \emph{effects of clutter}, which reduce parameter estimation accuracy, must be taken into account to assess sensing performance in ISAC systems properly. Clutter refers to unwanted echoes caused by static non-target objects in the background environment, which interfere with target reflections, obscure target parameters, and degrade sensing accuracy \cite{Skolnik2008}. To mitigate the clutter effect, \cite{Luo2024} proposed a static background clutter cancellation algorithm, along with a beamforming strategy to reduce interference between sensing and communication beams. A novel target detection algorithm for ISAC at mmWave frequencies, which models background clutter as multipath clusters, was in turn presented in \cite{Vinogradova2023}. Additionally, a background subtraction method for clutter removal was introduced in \cite{Ramos2024} to detect and track UE in cluttered indoor environments for sensing-aided communication systems. However, the background subtraction algorithm requires a reference signal that contains only the clutter channel, assuming that no other objects are present in the sensing area, which is critical for accurate parameter estimation.}

\textls[-1]{This paper proposes a two-dimensional rootMUSIC (2D-rootMUSIC) based clutter suppression algorithm with focus on the Third Generation Partnership Project (3GPP) bistatic ISAC scenario at mmWave frequencies, {where the clutter effect arises from signal scattering between the target and other background objects of no interest} \cite{Luo2024b,3GPP2024}. Unlike the prior-art background subtraction algorithm in \cite{Ramos2024}, the proposed method eliminates the need for a reference signal. The provided simulation results show that the proposed method enhances the signal-to-clutter-plus-noise ratio (SCNR) and accurately estimates target parameters under strong static background clutter, achieving performance comparable to the background subtraction algorithm. {Additionally, and importantly, the numerical results also show the superiority of the proposed method against prior-art when the reference signal measurement in the subtraction benchmark approach is subject to practical uncertainties.}}

\textit{Notations:} Italic lowercase letters $x$ denote scalar quantities. Bold lowercase letters $\mathbf{x}$ represent vectors, and bold uppercase letters $\mathbf{X}$ denote matrices. The set of complex-valued vectors or matrices with dimensions $a \times b$ is denoted by $\mathbb{C}^{a \times b}$. The operators $(\cdot)^T$ and $(\cdot)^H$ denote the transpose and Hermitian transpose, respectively. The term $e^x$ represents the exponential function. The notation $\mathcal{CN}(0, \sigma^2)$ denotes a circularly symmetric complex Gaussian distribution with zero mean and variance $\sigma^2$, while $\mathcal{N}(0, \sigma^2)$ denotes a real Gaussian distribution with the same parameters. The symbol $\mathbf{I}_N$ denotes an $N \times N$ identity matrix. The notation $\lvert \cdot \rvert$ represents the absolute value, $\lVert \cdot \rVert$ the norm, and $\mathbb{E}[\cdot]$ is the statistical expectation operator. The operator $\otimes$ represents the Kronecker product.

\iffalse
The rest of this paper is organized as follows. Section \ref{sec: Sys} presents the signal and channel models, followed by the proposed clutter suppression method. Section \ref{sec: Num} provides extensive numerical results to demonstrate the effectiveness of the proposed algorithm. Finally, Section \ref{sec: Conc} concludes the paper.
\fi

\section{System Model}
\label{sec: Sys}
\subsection{Fundamentals}
\begin{figure}[t!]
    \centering
    \includegraphics[width=0.95\columnwidth]{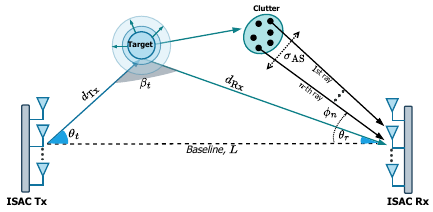}
    \vspace{-0.1cm}
    \caption{\textls[-4]{Bistatic sensing system model in a cluttered environment, following \cite{3GPP2024}.}}
    \vspace{-0.25cm}
    \label{fig:SysModel}
\end{figure}

\textls[-2]{A bistatic sensing scenario operating at mmWave is illustrated in \figref{fig:SysModel}, where the ISAC Tx is equipped with a uniform linear array (ULA) of $N_T$ antennas, and the ISAC Rx employs a ULA of $N_R$ antennas. The Tx is positioned at $\mathbf{p}_\mathrm{Tx} = [0,\,0]$, and the Rx is located at $\mathbf{p}_\mathrm{Rx} = [x_\mathrm{Rx},\,y_\mathrm{Rx}]$, resulting in a spatial separation with a baseline distance defined as $L = \lVert \mathbf{p}_\mathrm{Rx} - \mathbf{p}_\mathrm{Tx} \rVert$. In bistatic sensing mode, the Tx, target, and Rx form a bistatic triangle characterized by a bistatic angle $\beta_t = \pi - \theta_t - \theta_r > 0$, where $\theta_t$ denotes the azimuth angle of departure (AoD) and $\theta_r$ denotes the azimuth AoA. The case $\beta_t = 0$ corresponds to a monostatic sensing configuration. The distance from the Tx to the target is denoted by $d_\mathrm{Tx}$, whereas the distance from the target to the Rx is indicated by $d_\mathrm{Rx}$. Consequently, the bistatic distance is defined as $d_{\mathrm{Bis}} = d_\mathrm{Tx} + d_\mathrm{Rx}$, representing the total propagation path of the signal from the Tx to the target and subsequently to the Rx. The objective of the Rx is to detect the target and estimate its parameters by solving the bistatic triangle using the bistatic distance $d_{\mathrm{Bis}}$ and the AoA $\theta_r$. Since the ISAC Rx does not directly estimate $d_\mathrm{Tx}$ and $d_\mathrm{Rx}$, these distances can be inferred from geometric relationships using the known positions of the Tx and Rx, along with the estimated parameters $\theta_r$ and $d_\mathrm{Bis}$. The bistatic range relationships are given by}
\begin{align}
    &\frac{L}{\sin{\beta_t}} = \frac{d_\mathrm{Tx}}{\cos{\theta_r}} = \frac{d_\mathrm{Rx}}{\cos{\theta_t}} \label{eq: rng1},\\
    & L = \left({d}_\mathrm{Rx}^2 +  {d}_\mathrm{Tx}^2 - 2d_\mathrm{Tx}d_\mathrm{Rx}\cos{\beta_t}\right)^{1/2} \label{eq: rng2},\\
    &d_\mathrm{Rx} = \frac{d_{\text{Bis}}^2 - L^2}{2 (d_{\text{Bis}} - L\cos{\theta_r})} \label{eq: rng3}.
\end{align}

%Following the parameter estimation, the target position is given by $ \mathbf{p}_\mathrm{Tgt} = (x_\mathrm{Rx} - d_\mathrm{Rx}\cos{\theta_r}, \, y_\mathrm{Rx} + d_\mathrm{Rx}\sin{\theta_r})$.
In sensing systems, one of the most critical factors affecting the performance of target detection and parameter estimation is the bistatic radar cross section (RCS), which quantifies the amount of energy scattered by the target toward the receiver. Assuming a point-scatterer target model, the target is considered to reflect the incident signal uniformly in all directions with equal power, thereby behaving as an isotropic scatterer \cite{Skolnik2008}. In a single-target scenario, the received power at the Rx under free-space path loss is expressed as
\begin{equation}
    P_\mathrm{Rx} = \frac{P_\mathrm{Tx} G_\mathrm{Tx} G_\mathrm{Rx} \lambda_c^2 \alpha_{\mathrm{RCS},t}}{(4\pi)^3 d_\mathrm{Tx}^2 d_\mathrm{Rx}^2},
    \label{eq: P_Rx}
\end{equation}
where $P_\mathrm{Tx}$ is the transmit power, $\lambda_c$ denotes the wavelength, $\alpha_{\mathrm{RCS},t}$ denotes the bistatic RCS of the target, and $G_\mathrm{Tx}$ and $G_\mathrm{Rx}$ are the transmit and receive antenna gains, respectively.

It is worth noting that the point-scattering model holds when objects have a size comparable to $\lambda_c$ and have smooth surfaces \cite{Skolnik2008}. Objects with complex shapes and larger sizes, such as walls and buildings, may contain multiple scattering points distributed across the surface, all of which can backscatter the incident wave toward the Rx. The RCS of objects containing $N_{cl}$ scattering points can be given by \cite{Kell1965},
\begin{equation}
    \alpha_{\mathrm{RCS},c} = \left\lvert \sum_{n = 1}^{N_{cl}} \sqrt{\alpha_{n}}e^{j\psi_{n}} \right\rvert^2,
    \label{eq:RCS_Ncl}
\end{equation}
where $\alpha_{\mathrm{RCS},c}$ denotes the total RCS of the object, $\sqrt{\alpha_{n}}e^{j\psi_{n}}$ represents the complex reflectivity of the $n$-th scattering point, and $\alpha_{\mathrm{RCS},n} = \left\lvert \sqrt{\alpha_{n}}e^{j\psi_{n}} \right\rvert^2$ is the RCS of the $n$-th point on the object. 3GPP defines these types of objects as environmental objects (EOs Type-2) \cite{3GPP2024} that can cause clutter effects. The bistatic sensing scenario in a cluttered environment is demonstrated in \figref{fig:SysModel}. Herein, since the target reradiates power uniformly in all directions, the presence of clutter introduces non-line of sight (NLoS) paths, that is, the Tx-target-clutter-Rx paths, alongside the Tx-target-Rx path.
\subsection{Channel and Received Signal Models}
In mmWave systems, each multipath cluster corresponds to a physical object; therefore, clutter can be considered as a multipath cluster \cite{Vinogradova2023}. Consequently, since the ISAC Tx emits an OFDM waveform across $K$ active subcarriers and $M_s$ symbols to perform the sensing task, the target and clutter channel $\mathbf{H}_{k,m} = \mathbf{H}^t_{k,m} + \mathbf{H}^c_{k,m} \in \mathbb{C}^{N_R \times N_T}$ for the $k$-th subcarrier of the $m$-th OFDM symbol can be represented as
\begin{align}
    \mathbf{H}_{k,m} &= \mathbf{H}^t_{k,m} + \mathbf{H}^c_{k,m}
    =\sqrt{\alpha_{\mathrm{RCS},t}} e^{-j2\pi k\Delta f \tau_t} e^{j2\pi mT_s f_{D_t}}\nonumber \\
    &\times \mathbf{a}_\mathrm{Rx}(\theta_r)\mathbf{a}_\mathrm{Tx}^H(\theta_t) +  \sqrt{\frac{1}{N_{cl}}}\sum_{n = 1}^{N_{cl}}  \sqrt{\alpha_{\mathrm{RCS},t}  \alpha_{\mathrm{RCS},n}} \nonumber\\
    &\times e^{-j2\pi k\Delta f \tau_n} e^{j2\pi mT_s f_{D_c}} \mathbf{a}_\mathrm{Rx}(\theta_{c} + \phi_n)\mathbf{a}_\mathrm{Tx}^H(\theta_t), 
\end{align}
where $\Delta f$ is the subcarrier spacing, $T_s$ is the OFDM symbol duration, $\theta_t$ denotes the azimuth AoD, and $\tau_t$ and $\tau_n$ are the delays associated with the target and each ray in the clutter channel, respectively, with all clutter rays assumed to share the same delay $\tau_n$. The quantities $f_{D_t}$ and $f_{D_c}$ represent the Doppler shifts of the target and clutter, respectively. The angular spread in the clutter channel is modeled by representing the azimuth AoA of the $n$-th clutter ray as $\phi_n \sim \mathcal{N}(\theta_c, \sigma^2_\mathrm{AS})$, where $\theta_c$ is the mean azimuth AoA and $\sigma_\mathrm{AS}$ denotes the angular spread \cite{Hemadeh2018}. Although RCS is a deterministic quantity measured in square meters ($\mathrm{m}^2$), statistical approaches are commonly employed to model the behavior of objects with complex shapes. Thus, considering \eqref{eq:RCS_Ncl}, the clutter RCS can be modeled using the Swerling-I model \cite{Skolnik2008}, where $\alpha_{\mathrm{RCS},n} \sim \mathcal{CN}(0, \alpha_{\mathrm{RCS},c}/N_{cl})$ represents the RCS of the $n$-th ray in the clutter channel. Furthermore, since clutter results from the target isotropically scattering the signal, the total RCS contribution of the clutter is modeled as $\alpha_{\mathrm{RCS},t} \alpha_{\mathrm{RCS},n}$. The received array steering vector is given by $\mathbf{a}_\mathrm{Rx}(\theta_r) =  \begin{bmatrix}1, & e^{j\pi \sin\theta_r}, &\cdots, &e^{j\pi(N_R - 1)\sin\theta_r} \end{bmatrix}^T \in \mathbb{C}^{N_R \times 1}$, and the corresponding transmit steering vector $\mathbf{a}_\mathrm{Tx}(\theta_t) \in \mathbb{C}^{N_T \times 1}$ can be derived analogously.

The ISAC Rx collects or observes the backscattered signals from both the point-scatterer target and the static background clutter, which consists of \( N_{\mathrm{cl}} \) scattering points. The received signal thus includes echoes from both the target and the clutter channels. Therefore, the received signal for the $k$-th subcarrier of the $m$-th OFDM symbol is given by
\begin{equation}
    \mathbf{y}_{k,m} =  \left(\mathbf{H}^t_{k,m} + \mathbf{H}^c_{k,m}\right)\mathbf{w}_\mathrm{Tx} + \mathbf{z}_{k,m},
\end{equation}
\textls[-1]{where the involved beamformed transmit vector $\mathbf{w}_\mathrm{Tx}\,=\,\sqrt{{P_tG_t}/{N_T}}\mathbf{a}_\mathrm{Tx}(\theta_{n})x_{k,m} \in \mathbb{C}^{N_\mathrm{T} \times 1}$ embeds both the actual beamforming weights and the  transmit data symbols $x_{k,m}$ drawn from $M$-QAM constellation, while $\mathbf{z}_{k,m} \sim \mathcal{CN}(0, \sigma_N^2\mathbf{I}_{N_R}) \in \mathbb{C}^{N_R \times 1}$ denotes the vector of additive white Gaussian noise (AWGN) samples. Upon receiving the OFDM signal, assuming that the ISAC Tx and Rx share the information of transmitted symbols via a control unit, the element-wise division with $x_{k,m}$ is performed to remove their effect. Thus, $\mathbf{\hat{y}}_{k,m} = \mathbf{y}_{k,m}/x_{k,m}$ is a vector containing parameters to be estimated. }

In an OFDM frame, the Doppler shift corresponds to phase shifts across $M_s$ OFDM symbols, and the delay corresponds to phase shifts across $K$ OFDM subcarriers, which allow the Doppler and the delay to be considered as steering vectors in the slow and fast time domains, respectively \cite{Chen2023}. Consequently, the Doppler steering vector is given by
\begin{align}
   &\mathbf{d}({f}_D) =  \begin{bmatrix}
        1, & e^{j2\pi T_\mathrm{s}f_D}, &\cdots, &e^{j2\pi T_\mathrm{s}(M_s -1)f_D} 
    \end{bmatrix}^T,
   %&\mathbf{b}({\tau}) =  \begin{bmatrix}
        %1, & e^{-j2\pi \Delta_f\tau}, &\cdots, &e^{-j2\pi \Delta_f(K-1)\tau}
    %\end{bmatrix}^T,
\end{align}
where $\mathbf{d}({f}_D)\in \mathbb{C}^{M_s \times 1}$. Accordingly, the space-time vector can be defined as $\boldsymbol{\Psi}({f}_D,\theta) = \mathbf{d}({f}_D) \otimes \mathbf{a}_\mathrm{Rx}(\theta) \in \mathbb{C}^{M_sN_R \times 1}$. Thereafter, the received OFDM signal, after removing the transmit symbols, is expressed as follows:
\begin{align}
    \mathbf{\hat{y}}_k &= \mathbf{\hat{y}}_k^t + \mathbf{\hat{y}}_k^c  = \Big(\boldsymbol{\Psi}({f}_{D_t},\theta_r)\mathbf{a}_{\mathrm{Tx}}^H(\theta_t) e^{-j2\pi k\Delta f \tau_t} + \sqrt{\frac{1}{N_{cl}}}
   \nonumber \\
    & \times \sum_{n = 1}^{N_{cl}} \boldsymbol{\Psi}({f}_{D_c},\theta_c + \phi_n)\mathbf{a}_{\mathrm{Tx}}^H(\theta_t) 
   e^{-j2\pi k\Delta f \tau_c}\Big) \mathbf{\hat{w}}_\mathrm{Tx}
    + \mathbf{z}_k,
    \label{eq:ST_Y}
\end{align}
where $\hat{\mathbf{y}}_k^t$, $\hat{\mathbf{y}}_k^c$, and $\hat{\mathbf{y}}_k \in \mathbb{C}^{M_sN_R \times 1}$ denote the target, clutter, and total received signals for the $k$-th subcarrier, respectively. The term $\mathbf{z}_k \sim \mathcal{CN}(0, \sigma_N^2 \mathbf{I}_{M_sN_R}) \in \mathbb{C}^{M_sN_R \times 1}$ represents the vector of AWGN samples.
\section{Proposed Methods}
\subsection{Clutter Removal through Joint AoA/Doppler Estimation}
Since clutter represents an object of no interest, eliminating its effect is essential for accurate target parameter estimation. To address this, noise subspace-based algorithms are proposed for clutter suppression. Starting with the computation of the sample covariance matrix, the noise subspace is then obtained by performing eigenvalue decomposition (EVD) on the sample covariance matrix $\mathbf{R}_{\hat{\mathbf{y}}}$, yielding
\begin{equation}
    \mathbf{R}_\mathbf{\hat{y}} = \frac{1}{K}\sum_{k = 1}^K \mathbf{\hat{y}}_k\mathbf{\hat{y}}_k^H = \begin{bmatrix}
        \mathbf{U}_\mathbf{\hat{y}}^s \,\, \mathbf{U}_\mathbf{\hat{y}}^z 
    \end{bmatrix} \begin{bmatrix}
    \boldsymbol{\Sigma}_{\mathbf{\hat{y}}}^s & \mathbf{0} \\ \quad  \mathbf{0} & \boldsymbol{\Sigma}_{\mathbf{\hat{y}}}^z 
    \end{bmatrix} \begin{bmatrix}
        \mathbf{U}_\mathbf{\hat{y}}^s \,\, \mathbf{U}_\mathbf{\hat{y}}^z 
    \end{bmatrix}^H.
    \label{eq: cov_est}
\end{equation}

Let us further consider that the objects, which are target and clutter, in the sensing area are detected using the minimum description length (MDL) approach \cite{Wax1985}, and the number of detected objects is equal to $L$. Thus, in \eqref{eq: cov_est}, $\mathbf{U}_\mathbf{\hat{y}}^s$ yields a matrix with $L$ columns corresponding to the signal eigenvectors, and $\mathbf{U}_\mathbf{\hat{y}}^z$ denotes a matrix with $(M_sN_R - L)$ columns corresponding to the noise eigenvectors, which are utilized for parameter estimation through MUSIC or rootMUSIC algorithms. 

\textls[-2]{Therein, MUSIC and rootMUSIC algorithms are extensively employed for parameter estimation in ISAC systems \cite{Gonzalez2024, Sturm2011, Chen2023}. However, determining the AoA becomes challenging in the presence of clutter, as distinguishing the target's AoA from that of the clutter is not straightforward. The key difference between the target and clutter lies in their Doppler frequencies, with the Doppler frequency of the clutter being $f_{D_c} = 0$ due to its stationary nature. Consequently, the impact of clutter can be mitigated through joint angle-Doppler estimation, where the estimated AoAs corresponding to zero Doppler shift can be excluded. To achieve this, the 2D-MUSIC algorithm can be utilized. The 2D-MUSIC spectrum for joint angle/Doppler estimation is given by}
\begin{equation}
    P\left(f_\mathrm{D_\omega}, \theta_\omega\right) = \frac{1}{\boldsymbol{\Psi}(f_\mathrm{D_\omega},\theta_\omega)^H \mathbf{U}_{\mathbf{\hat{y}}_n}^z \mathbf{U}_{\mathbf{\hat{y}}_n}^{z^H} \boldsymbol{\Psi}(f_\mathrm{D_\omega},\theta_\omega)},
\end{equation}
where peaks in $P\left(f_\mathrm{D_\omega}, \theta_\omega\right)$ correspond to the jointly estimated AoAs $\hat{\theta}_\omega$ and Doppler frequencies $\hat{f}_\mathrm{D_\omega}$. However, the 2D-MUSIC algorithm suffers from high computational complexity due to an exhaustive grid search across possible AoAs and Doppler shifts. Therefore, we propose a 2D-rootMUSIC-based algorithm for clutter suppression. To begin with, let us define the matrix
\begin{equation}
 \boldsymbol{\Lambda} = \begin{bmatrix}
\boldsymbol{\Lambda}_{11} & \cdots & \boldsymbol{\Lambda}_{1M_s} \\
\vdots & \ddots & \vdots \\
\boldsymbol{\Lambda}_{M_s1} & \cdots & \boldsymbol{\Lambda}_{M_sM_s}
\end{bmatrix},
\label{eq: Lambda_Mat}
\end{equation}
where $\boldsymbol{\Lambda} = \mathbf{U}_{\mathbf{\hat{y}}_n}^z \mathbf{U}_{\mathbf{\hat{y}}_n}^{z^H}$, and $\boldsymbol{\Lambda}_{ij} \in \mathbb{C}^{N_R \times N_R}$ for $i,j \in \{1,\cdots, M_s\}$. Furthermore, the Doppler and array steering vectors are expressed as in \cite{Lee2019}, where the Doppler steering vector is given by  $\mathbf{d}(z_{f_{D_\omega}}) = [1 \quad z_{f_{D_\omega}} \quad \cdots \quad z_{f_{D_\omega}}^{M_s-1}]^T$,  and the receive array steering vector is  $\mathbf{a}_\mathrm{Rx}(z_{\theta_\omega}) = [1 \quad z_{\theta_\omega} \quad \cdots \quad z_{\theta_\omega}^{N_R-1}]^T$.  Here, $z_{f_{D_\omega}} = e^{j2\pi T_s f_{D_\omega}}$ and $z_{\theta_\omega} = e^{j\pi \sin{\theta_r}}$. Subsequently, to estimate the azimuth AoAs via \eqref{eq: Lambda_Mat}, the vector $\mathbf{c}_{\theta_\omega}$ is defined as
\begin{equation}
   \mathbf{c}_{\theta_\omega} = \sum_{l=-(N_R-1)}^{N_R-1} \Omega_l\left(\sum_{j,i=1}^{M_s} z_{\theta_\omega}^{j-i}\boldsymbol{\Lambda}_{ij}\right),
   \label{eq: C_theta}
\end{equation}
where the function $\Omega_l(\mathbf{X})$ denotes the sum of the $l$-th diagonal elements of matrix $\mathbf{X}$. From \eqref{eq: C_theta}, the roots $z^m_{\theta_\omega}$, where $\omega \in \{1, \hdots, L\}$, that are closest to and lie within the unit circle are obtained. The azimuth AoA estimates are then calculated as
\begin{equation}
    \hat{\theta}_\omega = \sin^{-1} \left(-\frac{\mathrm{arg}(z^m_{\theta_\omega})}{\pi}\right),
\end{equation}
assuming half-wavelength spacing $d=\lambda_c/2$ between the elements of the ULA. Leveraging the azimuth AoA estimates, the $(i,j)$-th element of the matrix $\mathbf{F}$ for Doppler estimation is given by
\begin{equation}
    \mathbf{F}_{ij} = \mathbf{a}_\mathrm{Rx}^{H}(\hat{\theta}_\omega) \left(\sum_{j,i=1}^{M_s} z_{f_{D_\omega}}^{j-i}\boldsymbol{\Lambda}_{ij}\right)\mathbf{a}_\mathrm{Rx}(\hat{\theta}_\omega).
    \label{eq:Doppler_mat}
\end{equation}

Afterwards, similar to \eqref{eq: C_theta}, the vector whose roots are within and closest to the unit circle is computed as follows:
\begin{equation}
   \mathbf{f}_{f_{D_\omega}} = \sum_{l=-(M_s-1)}^{M_s-1} \Omega_l\left(\mathbf{F}\right).
   \label{eq: f_Doppler}
\end{equation}
From \eqref{eq: f_Doppler}, the roots $z^m_{f_{D_\omega}}$ are obtained. Consequently, the bistatic velocity estimates are given by,
\begin{equation}
    \hat{v}_\omega =  -\frac{\mathrm{arg}(z^m_{f_{D_\omega}})\lambda_c}{2\pi T_s}.
\end{equation}
\vspace{0.04 in}

\textls[-2]{Ultimately, the estimated AoA and bistatic velocity pairs, denoted by $\{\hat{\theta}_\omega, \hat{v}_\omega\}$, are obtained. Since the clutter has zero (or near-zero) velocity, the corresponding estimates and their associated AoAs can be discarded. The target velocity is then identified as $\hat{v}_t = \arg \max_{\omega \in L} \, \lvert\hat{v}_\omega\rvert$, where $\hat{v}_t$ is the estimated target velocity, and its corresponding angle is denoted by $\hat{\theta}_r$. However, the AoA estimates may still be affected by strong clutter, which can reduce accuracy. To suppress this residual effect, the covariance matrix in~\eqref{eq: cov_est} is projected onto the clutter and noise subspace using the parameters $\{\hat{f}_{D_t} = \hat{v}_t/\lambda_c, \hat{\theta}_r\}$, as:}
\begin{equation}
    \widetilde{\mathbf{R}}_{\hat{\mathbf{y}}}^c = \mathbf{P}_t \mathbf{R}_{\hat{\mathbf{y}}} = \left(\mathbf{I}_{M_sN_R} - \hat{\boldsymbol{\Psi}}_t\left(\hat{\boldsymbol{\Psi}}_t^H \hat{\boldsymbol{\Psi}}_t\right)^{-1} \hat{\boldsymbol{\Psi}}_t^H\right)\mathbf{R}_{\hat{\mathbf{y}}}, 
    \label{eq: R_tilde}
\end{equation}
where $\hat{\boldsymbol{\Psi}}_t = \boldsymbol{\Psi}(\hat{f}_{D_t}, \hat{\theta}_r)$, and $\mathbf{P}_t$ is a projection matrix that projects onto the orthogonal complement of the subspace spanned by $\hat{\boldsymbol{\Psi}}_t$. Afterwards, the vector in~\eqref{eq: C_theta} is computed using $\widetilde{\mathbf{R}}_{\hat{\mathbf{y}}}^c$ to refine the target's azimuth AoA estimate.

\subsection{Space-Time Receive Filter and Range Estimation}

Based on the covariance matrix in \eqref{eq: R_tilde}, the space-time receive filter is next designed to suppress clutter while preserving the target signal. This leads to a minimum-variance distortionless response (MVDR) problem, and the corresponding receive filter can be expressed as
\begin{equation}
   \mathbf{u}_\mathrm{Rx} = {{\widetilde{\mathbf{R}}_{\hat{\mathbf{y}}}^{c^{-1}}} \hat{\boldsymbol{\Psi}}_t}\left({\hat{\boldsymbol{\Psi}}_t^H {\widetilde{\mathbf{R}}_{\hat{\mathbf{y}}}^{c^{-1}}} \hat{\boldsymbol{\Psi}}_t}\right)^{-1},
   \label{eq: u_R}
\end{equation}
where $\mathbf{u}_\mathrm{Rx}$ denotes the receive filter vector. The solution in \eqref{eq: u_R} minimizes $\mathbf{u}_\mathrm{Rx}^H \widetilde{\mathbf{R}}_{\hat{\mathbf{y}}}^c \mathbf{u}_\mathrm{Rx}$ while maintaining unity gain in the direction of $\boldsymbol{\Psi}(\hat{f}_{D_t}, \hat{\theta}_r)$. The output of the receive filter is subsequently used to compute the SCNR, defined as
\begin{align}
    \mathrm{SCNR} = \sum_{k=1}^{K} \frac{\mathbb{E}\left[\lvert
\mathbf{u}_\mathrm{Rx}^H \hat{\mathbf{y}}_k^t \rvert^2\right]}{\mathbb{E}\left[\lvert
\mathbf{u}_\mathrm{Rx}^H \hat{\mathbf{y}}_k^c + 
\mathbf{u}_\mathrm{Rx}^H \mathbf{z}_k \rvert^2\right]}.
\end{align}

Next, the output of the receive filter is employed for bistatic range estimation by applying an IFFT across the OFDM subcarriers, expressed as
\begin{equation}
    p(n_A) = \left\lvert\sum_{k = 0}^{N_A-1} \mathbf{u}_\mathrm{Rx}^H \hat{\mathbf{y}}_{n} e^{-j2\pi k n_A/N_A}\right\rvert^2,
    %= \left\lvert \mathrm{IFFT}\left\{ \mathbf{u}_\mathrm{Rx}^H \hat{\mathbf{y}}_n \right\} \right\rvert^2,
\end{equation}
where $N_A \geq K$ is the IFFT size used for bistatic range estimation. The estimated bistatic range is then given by
\begin{equation}
    \hat{d}_\mathrm{Bis} = \frac{\hat{n}_A c}{\Delta f N_A},
    \label{eq: Rng_Est}
\end{equation}
where $c$ is the speed of light and $n_A = 0, \dots, N_A - 1$. In \eqref{eq: Rng_Est}, the peak of $p(n_A)$ occurs when $\lceil \hat{n}_A = \hat{d}_\mathrm{Bis} \Delta f N_A / c \rceil$, where $\lceil\cdot\rceil$ is the ceil operator \cite{Sturm2011}.
\begin{table}[!t]
\vspace{0.2 in}
\centering
\caption{Numerical evaluation parameters}
\label{tab:SimParam}
\resizebox{0.55\columnwidth}{!}{%
\begin{tabular}{|c|c|}
\hline
\textbf{Parameter}                   & \textbf{Value} \\ \hline\hline
Carrier frequency, $f_c$ [GHz]                          & 28              \\ 
Subcarrier spacing, $\Delta f$ [kHz]                     & 120             \\ 
OFDM active subcarriers, $K$                                  & 792             \\ 
OFDM FFT \& IFFT size                                  & 1024             \\ 
OFDM symbols, $M_s$                                & 12              \\ 
Tx and Rx antennas, $N_T, N_R$                           & 12               \\ 
Target RCS, $\alpha_{\mathrm{RCS},t}$ [$\mathrm{m}^2$]    & 1               \\ 
Clutter RCS, $\alpha_{\mathrm{RCS},c}$ [$\mathrm{m}^2$]    & 2               \\ 
Noise power, $\sigma_N^2$ [dBm]                   & -80             \\ 
Clutter rays, $N_{cl}$                             & 6              \\ 
Angular spread, $\sigma_\mathrm{AS}$            & 3$^\circ$             \\ \hline
\end{tabular}%
}
\end{table}
\begin{figure}[!t]
    \centering
    \includegraphics[width=0.72\columnwidth]{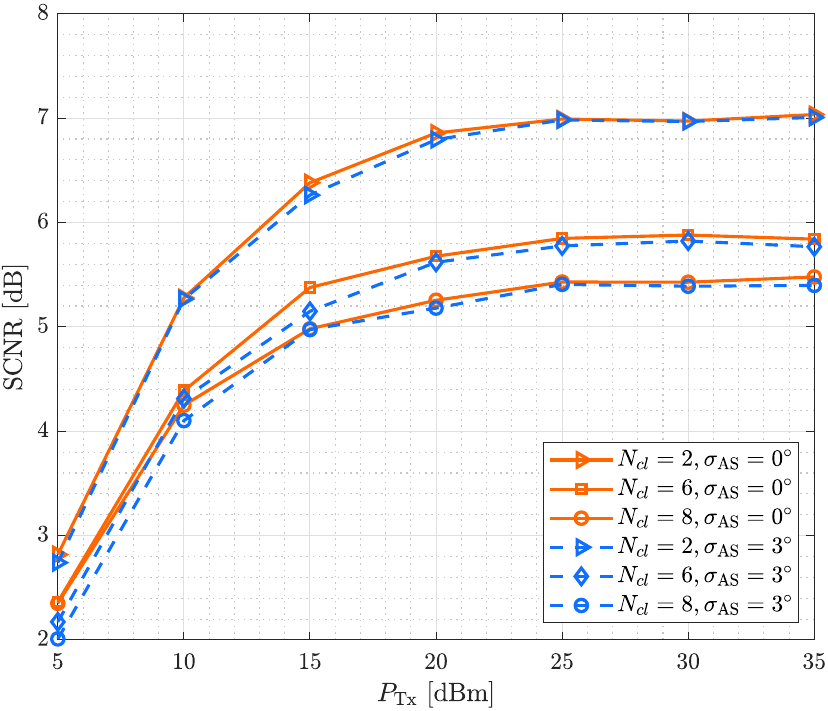}
    %\vspace{-0.2cm}
    \caption{SCNR for different values of $N_{cl}$ and $\sigma_\mathrm{AS}$ under varying $P_\mathrm{Tx}$.}
    \vspace{-0.25cm}
    \label{fig: SCNR}
\end{figure}
\section{Numerical Results}
\label{sec: Num}
\textls[-12]{In this section, extensive computer simulation results are presented to evaluate the effectiveness of the proposed scheme, while also benchmarking against prior-art. In the simulations, the target azimuth AoD $\theta_t$ and AoA $\theta_r$ are assumed to follow a uniform distribution $\mathcal{U}[20^\circ, 60^\circ]$, and target velocity $v_t$ follows \newpage $\mathcal{U}[10\,\mathrm{m/s}, 15\,\mathrm{m/s}]$, while the clutter AoA and velocity are set to $\{\theta_c= 10^\circ, v_c=0\,\mathrm{m/s}\}$. Since the clutter is illuminated by the signal scattered from the target, its AoD is assumed to be $\theta_t$. For simplicity, the Tx and Rx antenna gains are set to $G_\mathrm{Tx} = G_\mathrm{Rx} = 1$. The Rx is located at $\mathbf{p}_\mathrm{Rx} = [x_\mathrm{Rx} = 15\,\mathrm{m},\, y_\mathrm{Rx} = 0\,\mathrm{m}]$, and the target and clutter distances are computed using \eqref{eq: rng1}--\eqref{eq: rng3}, while the received powers corresponding to the target and each clutter ray are evaluated through \eqref{eq: P_Rx}. The number of active subcarriers is set to $K = 792$, corresponding to 66 physical resource blocks (PRBs) in 5G NR standard at FR2 with $\Delta f = 120$ \text{kHz} reflecting carrier bandwidth of $B = 100$ MHz \cite{3GPP5GNR}. Assuming a noise figure of 14 dB at the receiver, the noise power $\sigma_N^2$ is set to $-174+10\log_{10}(B)+14=-80$\,dBm. Unless otherwise specified, the parameters listed in Table~\ref{tab:SimParam} are used in the simulations, and the results are averaged over $10^3$ Monte Carlo simulations.}

The SCNR results for different $N_{cl}$ and $\sigma_\mathrm{AS}$ values are demonstrated for varying $P_\mathrm{Tx}$ values in \figref{fig: SCNR}. The SCNR improves as $P_\mathrm{Tx}$ increases across all cases and tends to saturate when $P_\mathrm{Tx}$ exceeds 25\,dBm. As the number of rays in the clutter channel increases, the SCNR decreases, indicating that the clutter signal becomes more dominant and interferes more with the target signal. Moreover, for all $N_{cl}$ values, only a slight performance degradation is observed when the angular spread increases from $\sigma_\mathrm{AS}=0^\circ$ to $\sigma_\mathrm{AS}=3^\circ$, validating the robustness of the proposed algorithm against angular spread introduced by the clutter channel.
\begin{figure}[!t]
\vspace{0.2 in}
    \centering
    \includegraphics[width=0.72\columnwidth]{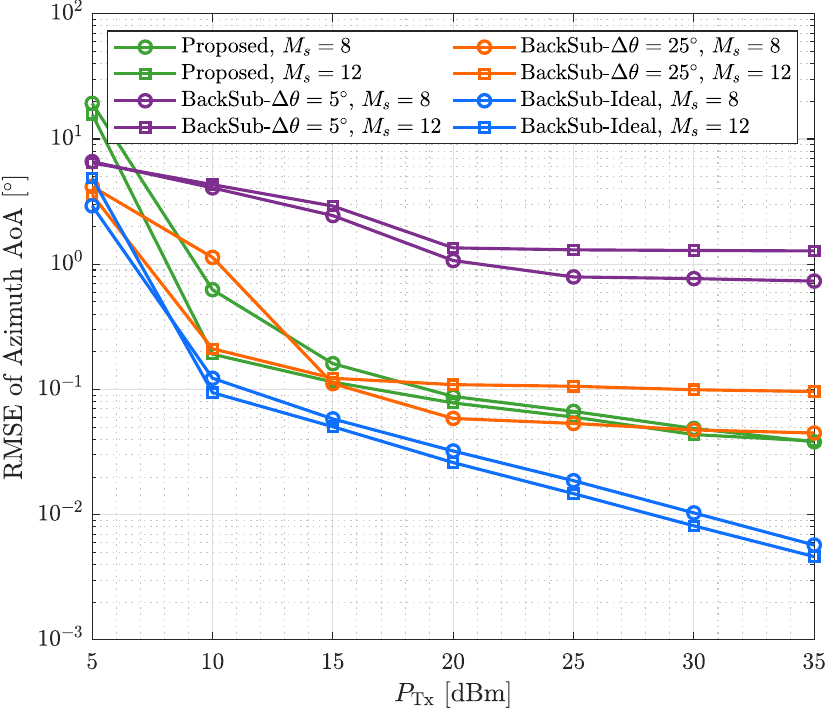}
    %\vspace{-0.2cm}
    \caption{RMSE of azimuth AoA estimation for the proposed method and two BackSub algorithm cases: a point-scatter object scenario with an AoA difference of $\Delta \theta$ from the target, and an ideal reference signal acquisition case, with varying $P_\mathrm{Tx}$ and different $M_s$ values.}
    \vspace{-0.25cm}
    \label{fig: RMSE_AoA}
\end{figure}
\begin{figure}[!t]
    \centering
    \includegraphics[width=0.72\columnwidth]{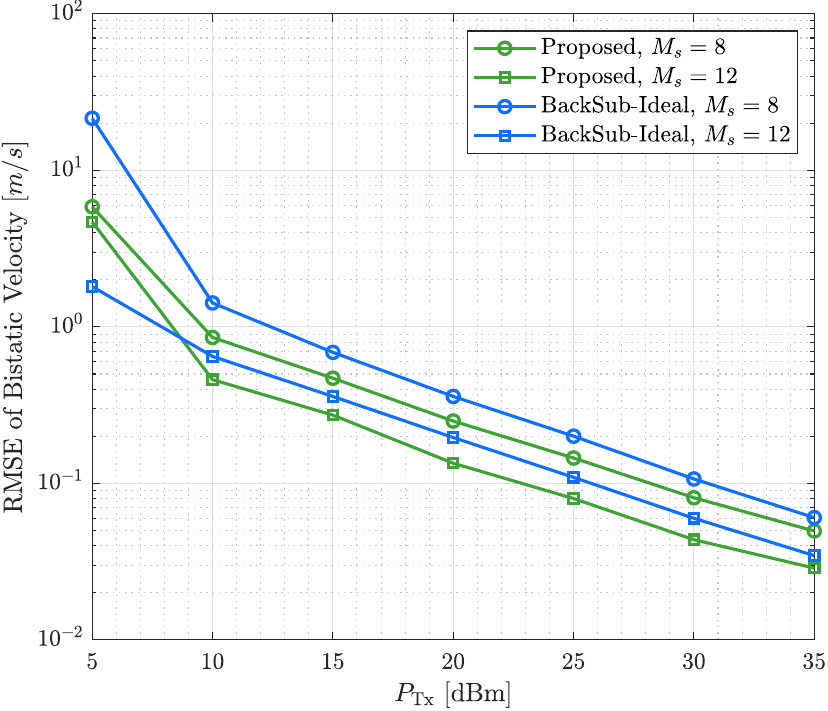}
    %\vspace{-0.2cm}
    \caption{RMSE of bistatic velocity estimation for the proposed method versus the ideal benchmark case of the BackSub algorithm, with varying $P_\mathrm{Tx}$ and different $M_s$ values.}
    \vspace{-0.25cm}
    \label{fig: RMSE_Vel}
\end{figure}

In Figs. \ref{fig: RMSE_AoA} and \ref{fig: RMSE_Vel}, the root mean square error (RMSE) performance of the proposed method is compared with the background subtraction (BackSub) benchmark algorithm presented in \cite{Ramos2024}, where 2D-rootMUSIC is employed for target parameter estimation. Ideally, the BackSub algorithm first captures a reference signal $\mathbf{y}_\mathrm{ref} = \mathbf{\hat{y}}_k^c\,+\,\mathbf{z}_k^\mathrm{ref}$ under the simplifying assumption that no actual target is present in the sensing area. When a target appears, the algorithm subtracts $\mathbf{y}_\mathrm{ref}$ from the received signal $\mathbf{\hat{y}}_k$ to remove the clutter effect. In contrast, the proposed method estimates target parameters directly from $\mathbf{\hat{y}}_k$, eliminating the need for a reference signal. As an additional more realistic baseline, \figref{fig: RMSE_AoA} considers also a scenario in which a point-scatterer object sharing the same RCS as the target, characterized by $\{\theta_x = \theta_r + \Delta \theta, v_x = 5\,\mathrm{m/s}\}$, is present during AoA reference signal acquisition. This yields $\mathbf{y}_\mathrm{ref} = \mathbf{\hat{y}}_k^x\,+\,\mathbf{\hat{y}}_k^c\,+\,\mathbf{z}_k^\mathrm{ref}$. It can be observed that the proposed method consistently outperforms the BackSub algorithm as transmit power $P_\mathrm{Tx}$ increases for $\Delta \theta = 5^\circ$. Although BackSub performance improves with greater azimuth AoA separation, especially at $M_s = 12$ and $\Delta \theta = 25^\circ$, the proposed method maintains superior performance at higher $P_\mathrm{Tx}$ values. Interestingly, increasing $M_s$ leads to worse performance in the more realistic evaluation case of the BackSub algorithm, as the increased contribution of clutter hinders effective suppression, while the presence of the point-scatterer object during reference signal acquisition further obscures the target parameters during estimation. In the ideal case, BackSub achieves better AoA estimation accuracy than the proposed method. 

As shown in Fig. \ref{fig: RMSE_Vel}, the proposed method achieves a lower RMSE in bistatic velocity estimation compared to the baseline, demonstrating robustness to strong background clutter. Better performance is observed for $M_s=12$ compared to $M_s=8$, since increasing the coherent processing interval (CPI) of the OFDM waveform, given by $M_s T_s$, enhances Doppler estimation accuracy. Furthermore, as the azimuth AoA and bistatic velocity are jointly estimated, extending the CPI also improves AoA estimation, as illustrated in Fig. \ref{fig: RMSE_AoA}.

\begin{figure}[!t]
    \centering
    \includegraphics[width=0.72\columnwidth]{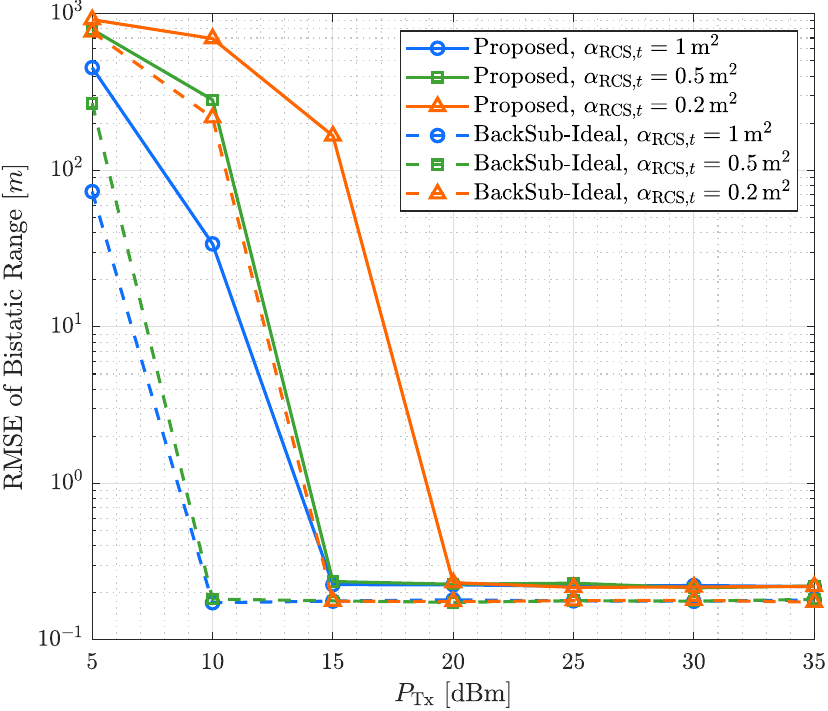}
        %\vspace{-0.2cm}
    \caption{RMSE of bistatic range estimation for the proposed method versus the ideal benchmark case of the BackSub algorithm, with varying $P_\mathrm{Tx}$ and different $\alpha_{\mathrm{RCS},t}$ values.}
    \vspace{-0.25cm}
    \label{fig: RMSE_Rng}
\end{figure}

Finally, the RMSE of bistatic range estimation is presented in \figref{fig: RMSE_Rng} for different target RCS values $\alpha_{\mathrm{RCS},t}$ as a function of varying $P_\mathrm{Tx}$ values. %, where the IFFT size is set to $N_A = 4K$, in \figref{fig: RMSE_Rng}. 
Predictably, performance degrades when $\alpha_{\mathrm{RCS},t} = 0.2\,\mathrm{m}^2$, as the clutter significantly masks the target parameters, limiting the effectiveness of the space-time filter in \eqref{eq: u_R}. However, as $\alpha_{\mathrm{RCS},t}$ increases, the proposed method accurately estimates the target parameters at lower $P_\mathrm{Tx}$ values, enabling the receive filter to suppress clutter more effectively and enhancing the accuracy of bistatic range estimation. Additionally, the RMSE results of the ideal BackSub algorithm are also included in \figref{fig: RMSE_Rng}. As expected, BackSub yields slightly better performance, as it assumes a target-only scenario in an ideal case by removing the background clutter channel. Notably, after $P_\mathrm{Tx} = 15\,\mathrm{dBm}$, and for $\alpha_{\mathrm{RCS},t}$ equal to 0.5 and 1\,$\mathrm{m}^2$, the estimation accuracy difference between the baseline and proposed methods is approximately 0.05\,m, underscoring that the proposed scheme can achieve comparable accuracy without relying on an additional reference signal.
\section{Conclusion}
\label{sec: Conc}
While larger array sizes and mmWave frequencies improve sensing accuracy in 6G ISAC systems, ensuring reliable performance in the presence of strong background ground clutter remains a challenge for perceptive networks. This paper presented a novel 2D-rootMUSIC-based clutter suppression algorithm tailored for bistatic ISAC systems. Simulation results have shown that the proposed method improves the SCNR, effectively suppresses strong background clutter, and enables accurate target parameter estimation, while achieving comparable results to the baseline algorithm without requiring a reference signal for clutter rejection.
\section*{Acknowledgement}
This work was funded by the EU Horizon Europe MSCA (MiFuture, Grant No. 101119643) on ultra-massive MIMO for future cell-free heterogeneous networks.

\bibliographystyle{IEEEtran}

\end{document}